\documentclass{aa}

\usepackage{graphics}

\begin{document}


   \title{Possible contribution of olefins and heteroatoms to the Unidentified Infrared Bands}
\author{R. Papoular
\inst{1}
          }

   \offprints{R. Papoular}

   \institute{Service d'Astrophysique and Service de Chimie Moleculaire, CEA Saclay, 91191 Gif-s-Yvette, France\\
              e-mail: papoular@wanadoo.fr
             }

   \date{ }
   \authorrunning {R. Papoular}
   \titlerunning {Contribution of olefins and heteroatoms to the UIBs}
   \maketitle
\begin{abstract}
The current assignments of the 11.3-$\mu$m feature are dicussed. An unbiased survey of correlation charts suggests that the olefinic group R$_{2}$C=CH$_{2}$ is a good alternative candidate. For the 12.7-$\mu$m feature, the best fits are provided by nitrites, R-O-N=O, and amines, R-N=H$_{2}$. Sulfones, SO$_{2}$, exhibit strong features near 7.7 and 8.6 $\mu$m, which may contribute to the UIBs. These additional functional groups are likely to be attached to the main hydrocarbon dust skeleton previously hypothesized.
 %
  \end{abstract}

%
\section{Introduction}

A large number of high-quality mid-IR spectra are now available in emission from HII regions (e.g. Cr\'et\'e  et al. \cite{cre99}), PDRs (PhotoDissociation Regions; see Bregman et al. \cite{breg94}), PNe (Planetary Nebulae; e.g. Allamandola et al. \cite{all89}, Molster et al. \cite{mol96}, Witteborn et al. \cite{ wit89}), transition objects (e.g. Buss et al. \cite{bus93}, Geballe et al. \cite{geb92}, Justtanont et al. \cite{jus96}, Kwok et al. \cite{kwo01}), M supergiants (Sylvester et al. \cite{syl94}), RNe (Reflection Nebulae; see Moutou et al. \cite{mou99}, Sloan et al. \cite{slo99}, Uchida et al. \cite{uch00}) and, in absorption, from protostars (e.g. Bregman et al. \cite{breg00}) and the GC (Galactic Center; see Chiar et al. \cite{chi02}, Tielens et al. \cite{tiel96}). Several  authors have assembled spectra of different origins (e.g. Boulanger and Cox \cite{boul98}, Peeters et al. \cite{pee02}) and given ``generic" (typical) spectra (e.g. Hony \cite{hon01}) of the UIBs (Unidentified IR Bands).

In the vast majority of these spectra, the ``11.3" feature peaks near 890 cm$^{-1}$ (11.25 $\mu$m) and emerges conspicuously from a plateau which extends from $\sim$11 to $\sim$ 13.5 $\mu$m. A second, clearly separate and typically weaker, feature also sits upon this plateau and peaks at 12.7  $\mu$m (790 cm$^{-1}$). When it is possible to delineate an underlying continuum for the plateau, the intensity of the latter is of the same order as that of the 12.7 feature, typically much weaker than that of the 11.3 feature. 

While the 12.7 feature remains unexplained, the 11.3 feature has been traditionally assigned to oop (out-of-plane) bending of C-H bonds at the periphery of aromatic clusters and in the absence of other, immediately adjacent, such bonds (solo bending). In laboratory spectra of PAHs (Polycyclic Aromatic Hydrocarbons) of different sizes and shapes, and various substitutions for peripheral hydrogen atoms, this type of vibration is found to occur indeed in the range 11-12 $\mu$m but, to my knowledge, no single known PAH displays a solitary, prominent, line at the peak position of the celestial feature, which is quite stable and well defined.  When a feature is present in a PAH, in the 11-12 $\mu$m range, it is generally accompanied by a few other bands of comparable intensity, near 12 and 13 $\mu$m; these are attributed to duo, trio and quartet bendings, i.e. \emph{coordinated} oop vibrations of, respectively, 2, 3 and 4 \emph{adjacent} C-H bonds (see Colthup et al. \cite{col90}, Smith \cite{smi79}). Present models are, therefore, unable to reproduce typical UIB spectra. This has recently prompted  reassessments of previous assignments, leading to various new proposals: essentially ionized PAHs (cations) and large, compact, neutral clusters. The present research addresses the same issue and suggests new assignments  for the 11.3 and 12.7 features, based on an unbiased perusal of IR data bases, extended to non-aromatics and to two heteroatoms among the most abundant in the universe: nitrogen and sulfur.

\section{Discussion of the current aromatic assignments of the 11.3 feature}
As implied by its designation, the solo C-H bending vibration is that of a C-H bond isolated from the next one by at least a bare C site. One conceivable way of obtaining this vibration without exciting duos, trios, etc. too is to take a linear chain of orthocondensed benzenic rings and bend it into a closed annulus: here, all the oop C-H bendings are indeed of the solo type. Unfortunately, such a structure is unknown in the laboratory and its solo vibration is not guarantied to fall at 11.25 $\mu$m.

Hony et al. (\cite{hon01}) proposed a solution consisting of a large diamond-shaped graphene (planar sheet made of compactly condensed hexagonal carbon rings). In this case, all C-H oop bends are again of the solo type, except at the four corners, where duos occur. Both objections raised above against the annular chain of benzene rings also apply here. Moreover, large graphene sheets are known to curl up and fold over into fullerenes or nanotubes (Robertson et al. \cite{rob92}), whose IR features were never observed in the sky. Finally, even if such large graphenes existed, they would necessarily display a strong  continuum because of their large electronic conductivity, which would intolerably reduce the UIB feature contrast.

For any known isolated PAH, enough dehydrogenation may conceivably eliminate duos, trios, etc. But, for one thing, there is no reason for dehydrogenation to remain nicely distributed around the cluster, so as to avoid all adjacencies. More importantly, the relatively high density of atomic hydrogen in UIB sources in general, and the high measured rate of H$_{2}$ formation on dust (R$\sim$3 10$^{-17}$cm$^{3}$sec$^{-1}$ in diffuse clouds) make it highly improbable that many dangling C bonds will remain simultaneously unoccupied for any long period of time.

For want of an ideal, \emph{single}, aromatic hydrocarbon which would mimic the single 11.3 feature, let us now explore the possibility that aromatic hydrocarbons \emph{as a family} could mimic the whole, or part, of the 11-14 $\mu$m spectrum. For this purpose, Papoular et al. (\cite{rp89}) displayed the average of the spectra of the 18 smaller PAHs up to coronene. As expected, this composite spectrum exhibits a ``forest" of features, between 11 and 14 $\mu$m, which blend into an asymmetrical band peaking at $\sim$13.7 $\mu$m, in sharp contrast with the UIB spectrum.

Later on, Schlemmer et al. (\cite{schl94}) studied the emission spectra of ``hot" gaseous molecules of naphtalene and pyrene between 3 and 7.5 $\mu$m. They noted the presence of strong features between 5 and 6 $\mu$m (not observed in the sky) and a ratio of mid- to near-IR feature intensities much lower than in UIB spectra. They concluded that small, neutral, PAHs cannot be the carriers of the UIBs. Nevertheless, Allamandola and co-workers (see references in Allamandola et al. \cite{all99}) extended the study of neutrals to larger sizes, as far as dicoronylene C$_{43}$H$_{20}$, isolated in low-temperature rare gas matrices. Spectra of individual, and mixtures of, neutral PAHs are given in Allamandola et al. (\cite{all99}), Hudgins et al. (\cite{hud99}), Sloan et al. (\cite{slo99}) and Hony et al. (\cite{hon01}). These results confirm the expectation, in the spectra of neutral PAHs, of 4 dominant oop C-H bending bands, peaking around 11.3 (solo), 12 (duo), 13 (trio) and 13.5 $\mu$m (quartet). All intensities are of the same order, so no mixture could be proposed to mimic the prominent, isolated, 11.3 feature of generic UIB spectra. 

The same can be said of the random mixtures of small and intermediate PAHs which are known to constitute the backbone of bulk kerogens (Papoular  \cite{rp01}) and coals (Papoular et al. \cite{rp89}).\emph{ For these materials, the oop C-H bending bands form a plateau extending from $\sim$11 to $\sim$13.5 $\mu$m, with 3 or 4 low-contrast peaks corresponding to the various adjacencies}. Guillois et al. (\cite{gui96}) showed that, as a consequence, these materials correctly mimick the rippled 12-$\mu$m plateau of type-B sources (see Tokunaga \cite{tok97} and Kwok et al. \cite{kwo01}). This is not the case with type-A objects, where the plateau is also present, but dominated by a lone and narrower 11.25-$\mu$m feature in which we are primarily interested here. 

Theorists in the early '90s started \emph{ab initio} spectral calculations of PAHs, including charged ones, in the hope of achieving a better fit to the observed relative feature intensities. Langhoff (\cite{lang96}) then produced high-quality theoretical spectra of PAH neutrals, cations and anions, using density functional theory. This demonstrated once again the inadequacy of neutrals as UIB carriers, but showed that cations do a far better job  regarding relative feature intensities, although the frequency agreement with observations was worsened to some extent. The 11.3 feature was tentatively assigned to the duo C-H oop mode, and the 12.7 to the trio mode of \emph{cations}. 

These results fuelled the interest of experimentalists and, a few years later, a rich laboratory data base was produced (see Sloan et al. \cite{slo99} and Hony et al. \cite{hon01} for references and typical spectra). From these rather concurrent theoretical and experimental results, it appears that, in the spectral range of interest here,

--the solo oop C-H bending of neutrals fits the celestial 11.3 feature in frequency far better than does that of cations; but it is too strong;

--the solo oop C-H bending of cations falls near 11$\mu$m, and so may be the carrier of the weak feature often detected on the blue wing of the 11.3 feature; but if this were the case, and if cations were ubiquitous, the 11 $\mu$m feature should be much stronger than observed;

--both neutrals and cations also exhibit equally strong duos, trios, etc. as far as 13.5 $\mu$m; so neither can explain the dominance of the 11.3 feature;

--both in theory (see Langhoff \cite{lang96}; Bakes et al. \cite{ba01}) and in experiment (see Hony et al. \cite{hon01}), the whole spectrum of PAHs depends heavily on their degree of ionization; one would therefore expect very different spectra from regions of different UV field intensities as measured by their G values. In fact, the observed UIB spectra vary only slightly from G=1 to G=10$^{4}$ (see Boulanger and Cox \cite{boul98}; Sellgren \cite{sel01}); it seems therefore that, in the framework of the PAH model of the UIBs, this remakable stability can hardly be accounted for, unless one is prepared to invoke very stringent and finely tuned electrical, optical and thermal properties for these molecules, as well as very well defined mixtures thereof, depending on environmental conditions, the reality or realizability of which have yet to be proven (see Verstraete et al. \cite{vers01}; Bakes et al. \cite{ba01}). We have therefore to conclude with Boulanger and Cox that our problem is not likely to be solved by ionized PAHs.

Even the loose assignment of the 11.3 feature to the general class of aromatic bonds is called into question by angular scans of the sky, indicating the non-coincidence of the intensity peak of this feature with that of the definitely aromatic C-H stretch feature at 3.28 $\mu$m (Bregman et al. \cite{breg94}; Cr\'et\'e et al. \cite{cre99}).

\section{Non-aromatic assignment of the 11.3-$\mu$m feature}
The slight variations in band shape and peak wavenumber of the UIBs from different sources suggest that these bands are mostly carried by \emph{functional groups} rather than particular molecules with definite spectra. These groups are made of small numbers of atoms, always bonded together in a configuration whose geometry changes only slightly according to the main structure to which it is attached. Each group is characterized by a small number of vibration modes. \emph{Correlation charts have been drawn, which graphically display the vibration strength and the range covered by the changing peak wave number of each mode of every group as its attachment changes}. These charts are not limited to aromatic hydrocarbon groups, but include olefinic and aliphatic bondings as well as higher mass atoms. No group is expected to carry all the UIBs. We have surveyed such charts in Colthup et al. (\cite{col90}), Smith (\cite{smi79}) and others, in search of alternative assignments to the 11.3 feature, then to the 12.7 feature (see Sec. 4). 

The best fit to the former turns out to be the (olefinic) \emph{vinylidene} group R$_{2}$C=CH$_{2}$, where R represents an alkane substituent (saturated carbon; see fig. 1). This has a \emph{strong} CH$_{2}$ wag in the range 885-895 cm$^{-1}$ (11.2-11.3 $\mu$m), in which both H atoms oscillate in phase and oop. The other IR-active modes are: a CH$_{2}$ deformation (weak) near 1415 cm$^{-1}$ (7.07 $\mu$m); a C=C stretch (medium strong) at 1639-1661 cm$^{-1}$ (6.02-6.1 $\mu$m); a weak overtone of the CH$_{2}$ wag at 1775-1792 cm$^{-1}$ (5.58-5.63 $\mu$m); a weak symmetric (in phase) stretch near 2985 cm$^{-1}$ (3.35 $\mu$m) and a medium strong asymmetric (out-of-phase) C-H stretch at 3070-3100 cm$^{-1}$ (3.23-3.26 $\mu$m). An example of known molecule including this group is dimethylethylene, where R=CH$_{3}$. The vinylidene group may also be attached to a 5- or 6-membered carbon ring without noticeable spectral changes, as in methylene cyclohexane  (Colthup et al. \cite{col90}).
\begin{figure}
\resizebox{\hsize}{!}{\includegraphics{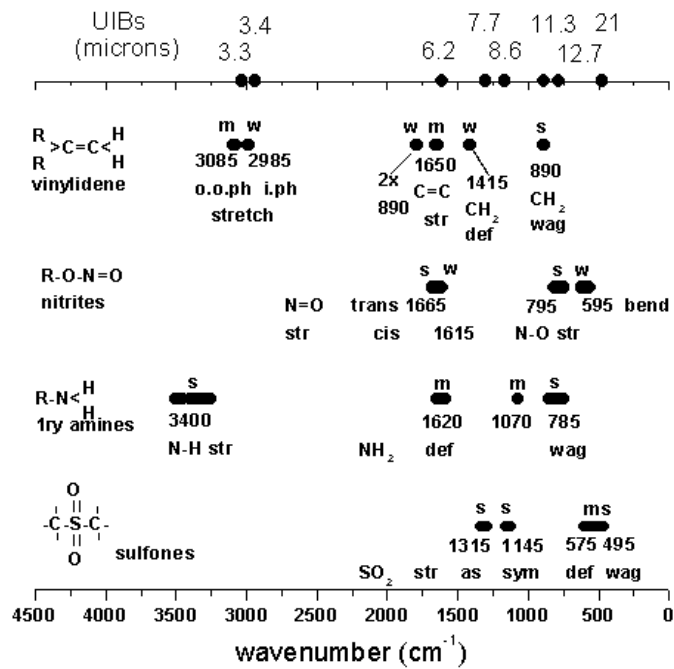}}
\caption[]{Partial correlation chart. R: alcane substituant. Feature strength: w,weak; m, medium; ms, medium-strong; s, strong. Modes: str, stretch; def, deformation; iph, in phase; ooph, out of phase.Numbers are for band centers.}
\end{figure}
Arguments in favour of this assignment are: a) there are no misplaced strong features and no added continuum; b) two secondary features fall in the vicinity of weak UIBs observed by Peeters et al. (\cite{pee02}) at 6 and 7 $\mu$m; c) the shift of the asymmetric CH stretch relative to the 3.28-$\mu$m feature may help explaining the large variations observed in the ratio of the 3.3 to 11.3 UIBs (see Bregman et al. \cite{breg94}) and the lack of spatial coincidence of the peak intensities of these bands (Cr\'et\'e et al. \cite{cre99}); d) the non-aromaticity of the group may explain part of the large fluctuations of the intensity ratio of the 6.2 to 11.3 UIBs (see Hony et al. \cite{hon01}); e) this assignment is consisitent with the established presence in space of carbon double bonds as in cumulene carbenes (see Thaddeus et al. \cite{tha98}).

In the similar \emph{vinyl} group, one of the R's of vinylidene is replaced by an H (Colthup et al. \cite{col90}). This is interesting in that the strong CH$_{2}$ wag is then shifted to 905-910 cm$^{-1}$ ($\sim$11 $\mu$m), near the weak feature observed in the wing of the 11.25-$\mu$m UIB (see Sloan et al. \cite{slo99}). Vinyl also has a (weaker) \emph{trans} C-H wag at 985-995 cm$^{-1}$ ($\sim$10.1 $\mu$m), in which the 2 opposite C-H bonds oscillate oop. If vinyl groups are present in space, this feature should be detectable in the UIB spectrum whenever the 11.1 feature is observed, as is perhaps the case in Sloan et al. (\cite{slo99}).

\section{The 12.7-$\mu$m feature ($\sim$787 cm$^{-1}$)}
In space, the 12.7 feature is not directly correlated with the 11.3 UIB (see Hony et al. \cite{hon01}) and, indeed the candidates considered above do not exhibit this feature. An independent search is therefore required for it. This feature falls in the gap between the duo and trio aromatic C-H oop peaks. Exceptionally, naphtalene does exhibit a strong feature at 780 cm$^{-1}$ (see Schrader \cite{schr89}); numerical simulation of this molecule (Papoular \cite{rp01b}) shows that this frequency corresponds to all 8 CH's wagging in phase oop, the two opposite edge pairs being most active.

A somewhat similar geometry is provided by para-xylene, a benzene ring where CH$_{3}$'s are substituted for 2 opposite H's , which carries a very strong feature at 793 cm$^{-1}$ (Schrader \cite{schr89}). Several other para-substituted benzenes exhibit a very strong feature in the range covered by the 12.7 UIB. 4-picoline C$_{6}$H$_{7}$N has a closely related symmetry (Schrader \cite{schr89}): one C is substituted with N and the opposite H is replaced by a methyl (CH$_{3}$); the resonance at 799 cm$^{-1}$ again corresponds to 2 opposite, in phase, duo C-H wags and is very strong. An interesting feature of these molecules is their strong vibrations near 21 $\mu$m. On the other hand, they also have undesirable, albeit weaker, bands near 7 $\mu$m. 

It seems, therefore, preferable to turn to primary and secondary \emph{organic nitrites} R-O-N=O, of which there are two isomers, according to the relative orientations of the R-O and N=O bonds (Lin-Vien et al. \cite{lin90}). The desired vibration is the \emph{strong} N-O stretch at 775-814 cm$^{-1}$ (12.3-12.9 $\mu$m; fig. 1). There is only one other main band, due to N=O stretching, at 1613-1681 cm$^{-1}$ and likely peaking near 6$\mu$m. A subsidiary band occurs at 565-625 cm$^{-1}$, due to group bending and is reminiscent of the weak 16.4 $\mu$m-band observed by Moutou et al. (\cite{mou00}).

In aliphatic \emph{amines}, R-NH$_{2}$, the NH$_{2}$ wagging and twisting vibrations give rise to broad, strong, usually multiple bands around 800 cm$^{-1}$  (Colthup et al. \cite{col90}), and so may also be considered as a candidate. A test of its presence could be the detection of its N-H stretch, which is solitary around 3400 cm$^{-1}$ (fig. 1), but hard to observe because of telluric absorption. 

\section{Inclusion in condensed dust}
As a by-product of the present correlation search, \emph{sulfones}, SO$_{2}$ turned out to have 2 strong stretch bands near the 7.7- and 8.6-$\mu$m UIBs, as well as weaker deformation around 17.4 and wag around 20 $\mu$m (fig. 1). Since O, N and S are among the most abundant elements, and since molecular combinations thereof have been detected in space, the classes of functional groups in fig.1 may all be considered as possible contributors to the UIBs. The shape and variability of the latter speak in favour of the groups being included randomly in condensed matter rather than molecules. A possible model of that matter is provided by the 3-D models of natural coals and kerogens, which invoke an amorphous hydrocarbon, only partly aromatic, interspersed with substitutions for hydrogen and carbon atoms with precisely the heteroatoms considered above: O, N, S (see Papoular et al. \cite{rp89} and Papoular \cite{rp01}), and ``a complex collection of alkane and alkene side chains" (Kwok et al. \cite{kwo01}). In this framework, the coal/kerogen-like material, in its less evolved forms, provides models for the broad, mostly aliphatic, features of B-type objects. 

On the other hand, the large range of its more mature (aromatic) forms, augmented with adequate concentrations of functional groups such as those proposed above, will help tailoring the great variety of relative intensities, bandwidths and contrast of aliphatic and aromatic features in celestial spectra (see Peeters et al. \cite{pee02}), with no need to invoke dehydrogenation, ionization or size effects. By analogy with laboratory heat treatment (Papoular et al. \cite{rp89}), this continuum of maturation states (tending towards type-A objects) may result, in space, from successive gas shocks or radiative heating or, still, photochemistry (Kwok et al. \cite{kwo01}).

\end{document}